\documentclass{ws-ijmpa-blank-new}
\usepackage[super,compress]{cite}
\usepackage{graphicx}

\usepackage{amssymb,epsf,latexsym,amssymb,amsmath,mathrsfs}
\usepackage[english]{babel}

\newcommand{\beq}{\begin{equation}}
\newcommand{\eeq}{\end{equation}}
\newcommand{\beqa}{\begin{eqnarray}}
\newcommand{\eeqa}{\end{eqnarray}}
\newcommand{\bsubeqs}{\begin{subequations}}
\newcommand{\esubeqs}{\end{subequations}}

\makeatletter
\@addtoreset{equation}{section}
\renewcommand{\theequation}{\thesection.\arabic{equation}}
\makeatother

\begin{document}
\markboth{F.R. Klinkhamer}
{Ultraheavy particles at the LHC or a next-generation collider?}

%
\catchline{}{}{}{}{}
%

\title{\vspace*{-9mm}
Ultraheavy particles at the LHC or a next-generation collider?}

\author{F.R. KLINKHAMER}

\address{Institute for Theoretical Physics, Karlsruhe Institute of
Technology (KIT),\\ 76128 Karlsruhe, Germany\\
frans.klinkhamer@kit.edu}

\maketitle


\begin{abstract}
\\
If the effective cosmological constant $\Lambda$ of the present universe
is due to physical processes in the early universe
operating at temperatures just above the electroweak energy scale,
it is possible that new particles with multi--$\text{TeV}$ masses exist.
These ultraheavy particles may (or may not) show up at
the Large Hadron Collider (LHC) or  a next-generation proton-proton collider. If they do, they may provide new insights into
the early universe and fundamental physics.
\\
\end{abstract}
\vspace*{.00005\baselineskip}
\hspace*{0mm}
{\footnotesize
\emph{Journal}:
\emph{Mod. Phys. Lett. A} \textbf{30}, 1550149 (2015)
\vspace*{.25\baselineskip}\newline
\hspace*{5mm} \emph{Preprint}: arXiv:1503.03858
}
\vspace*{-5mm}\newline
\keywords{models beyond the standard model,
dark energy,
general relativity,
cosmology}
\ccode{PACS No.:14.80.-j}


\section{Introduction}
\label{sec:Introduction}

In a series of papers,\cite{KV2009-ew-kick,K2010-Lambda-TeV,%
K2011-Lambda-TeV-simple-model} we have argued that the
effective cosmological constant $\Lambda$ of the present universe,
interpreted as a remnant vacuum energy density,
may be due to the imprint of ultraheavy particles on the
Hubble expansion of the early universe.
The discussion is in the framework of the so-called $q$-theory
which describes the evolution of the \emph{macroscopic} gravitating
vacuum energy density $\rho_{V}[q]$ due to a microscopic conserved
relativistic vacuum variable $q$  
(the original papers are Refs.~\citen{KV2008-statics,KV2008-dynamics}
and a one-page review appears as App.~A in Ref.~\citen{KV2011-review}).
In view of the upcoming Run 2 of
CERN's Large Hadron Collider (LHC), it may be
timely to review the main assumptions of the
argument and to clarify the meaning of the predictions.
Natural units will be used with $\hbar=1$ and $c=1$.

At this point, it may already be worthwhile to present the basic
equation for the remnant vacuum energy density
in a flat Friedmann--Robertson--Walker (FRW) universe
with cosmic time $t$,
\beq\label{eq:Lambda-theory}
\Lambda
\equiv \lim_{t\to\infty}\;\rho_V(t)
=
r_{V\infty}
\;M^{8}\, /(E_{P})^{4} \,,
\eeq
where
$r_{V\infty}$ is a nonnegative number,
$M$ the mass scale of the hypothetical new particles,
and $E_{P}$ the reduced Planck energy,
\beq\label{eq:E-planck-def}
E_{P}
\equiv
\sqrt{1/(8\pi G_{N})}
\approx
2.44\times 10^{18}\:\text{GeV}.
\eeq
Inverting \eqref{eq:Lambda-theory} gives
the following expression for the new mass scale $M$:
\beqa\label{eq:M-estimate}  
\hspace*{-10mm}
M
&=&
(r_{V\infty})^{-1/8}\,\Lambda^{1/8}\, (E_{P})^{1/2}
\approx
5.56\;\text{TeV}
\left(\, \frac{10^{-3}}{r_{V\infty}}\, \right)^{1/8}
\left(\, \frac{\Lambda^{1/4}}{2.25\;\text{meV}} \,\right)^{1/2},
\eeqa
where the numerical value used for $\Lambda$
follows from  Table~2.1 in Ref.~\citen{PDG2014}.

Taking Eq.~\eqref{eq:Lambda-theory} as it stands, the number $r_{V\infty}$
can be interpreted as an efficiency factor
for producing a remnant vacuum energy density
given the energy scales involved, $M$ and $E_{P}$.   
Remark that the parametric dependence of \eqref{eq:Lambda-theory}
has already been discussed
by Arkani-Hamed \textit{et al.},\cite{ArkaniHamed-etal2000}
but without a convincing theory of
how the vacuum energy density evolves (it is here that
$q$-theory\cite{KV2008-statics,KV2008-dynamics} is supposed to take over).
Still, the authors of Ref.~\citen{ArkaniHamed-etal2000}
discuss persuasively the role of \eqref{eq:Lambda-theory} for the
so-called  triple cosmic coincidence puzzle:
why are the orders of magnitude of the energy densities of vacuum, matter,
and radiation approximately the same in the present Universe?

Taking the point of view that
the effective cosmological constant $\Lambda$ of the
present universe has been measured,\cite{PDG2014}
Eq.~\eqref{eq:M-estimate}
can be read as a prediction of the mass scale $M$, provided
the ``efficiency factor'' $r_{V\infty}$ is known.

The task, then, is to \emph{calculate}
the pure number $r_{V\infty}$ entering \eqref{eq:Lambda-theory}.
This calculation is extremely difficult and, up till now, there is
only an approximate phenomenological
description available.\cite{KV2009-ew-kick,K2010-Lambda-TeV,%
K2011-Lambda-TeV-simple-model}
In the present paper, we try to simplify the discussion as
much as possible, in order to highlight the crucial assumptions
of the argument.  For definiteness, we assume $q$ to come from
the field strength of a three-form gauge field,\cite{KV2008-statics}
so that $q$ has mass dimension 2.

\section{Setup}
\label{sec:Setup}

\subsection{$\boldsymbol{K}$--freezing model}
\label{subsec:K-freezing-model}

In the framework of $q$-theory,\cite{KV2008-statics,KV2008-dynamics,KV2011-review}
the analysis of  Ref.~\citen{KV2009-ew-kick} has shown that
the sudden presence of ultraheavy particles (possibly created by a phase transition
with decay afterwards) perturbs the Hubble expansion and kicks the
vacuum energy density $\rho_{V}(t)$ away from zero to a small positive value. 
This process occurs at a cosmic age of order
\beq\label{eq:t-kick-def}
t_\text{kick}  \equiv  E_{P}/M^{2} = \xi^{1/4}\;M^{-1}\,,
\eeq
where the last expression has been
written in terms of the energy-density hierarchy parameter
\beq\label{eq:xi-def}
\xi\equiv  \big(E_{P}/M \big)^4\,.
\eeq
In the same way as the authors of Ref.~\citen{ArkaniHamed-etal2000},
we initially do not worry about what physics stabilizes the large hierarchy
between $M \sim \text{TeV}$ and $E_{P}\sim 10^{15}\;\text{TeV}$.
In other words, we leave aside the well-known hierarchy problem
and simply try to determine the mass scale $M$  from cosmology,
for the given value of $E_{P}$.

The fundamental issue is the freezing of the vacuum energy
density created by the kick.
As realized in Ref.~\citen{K2011-Lambda-TeV-simple-model},
this freezing can be modeled by
a time-dependent (or, better, temperature-dependent)
gravitational coupling; see also Sec.~\ref{subsec:Dissipation-model}
for further comments.
Specifically, we take a Brans--Dicke-type term in the action density,
\beq\label{eq:EH-term}
\mathcal{L}_\text{grav} = K[q,\Phi]\,R[g] \,,
\eeq
where $\Phi$ stands for one or more of the new matter fields.
Recall that the standard Einstein theory has an action-density
term $K_0\,R[g]$  with constant $K_0=1/(16\pi G_N)$.
In a flat FRW universe with cosmic time $t$,
we now assume the following simplified behavior:
\bsubeqs\label{eq:KAnsatz-all}
\beqa\label{eq:KAnsatz}
K[q,t]
&=&
q(t)/2 +\theta_{K}(t)\; \big[q_0/2-q(t)/2\big]\,,
\\[2mm]
\label{eq:KAnsatz-thetaK}
\theta_{K}(t) &=& \theta(t-t_{K})\,,
\eeqa
\esubeqs
in terms of the standard  stepfunction
\beqa\label{eq:KAnsatz-theta-def}
\theta(t)&=&
\left\{\begin{array}{l}
  1   \;\;\;\; \text{for}\;\;\;\;  t >    0  \,, \\[0mm]
  0   \;\;\;\; \text{for}\;\;\;\;  t \leq 0 \,.
\end{array}\right.
\eeqa
Observe that $q_0$ in \eqref{eq:KAnsatz} corresponds to the
constant equilibrium value of the vacuum variable $q$ and that
the inverse of $q_0$ gives Newton's constant, $G_N = 1/(8\pi q_0)$.

We also let the coupling constant of the ultraheavy matter component
be controlled by another stepfunction,
\bsubeqs\label{eq:gAnsatz-all}
\beqa\label{eq:gAnsatz}
g^{2}(t)
&=&
\overline{g}^{2}\;\theta_{g}(t)\,,
\\[2mm]
\label{eq:gAnsatz-thetag}
\theta_{g}(t) &=& \theta(t-t_{g})\,.
\eeqa
\esubeqs
This time-dependent coupling constant generates an ultraheavy
matter component, even if it is not present initially ($t<t_{g}$).
At a later moment, the ultraheavy particles decay,
which can be modelled by letting $g^{2}(t)$ drop to zero again
and by having a constant decay constant $\lambda^{2}$ instead.
For the kick mechanism to operate (that is, $r_{V}$  first
kicked away from zero and then frozen), the following inequality is
required:
\beq\label{eq:tg-tM-inequality}
t_{g} < t_{K} \,.
\eeq
Moreover, the kick mechanism only makes sense if
these two timescales are of the same order of magnitude
(see Sec.~\ref{sec:Conclusion} for further discussion).

Next, define dimensionless variables\cite{K2011-Lambda-TeV-simple-model}
by use of the energy scales $M$ and $E_{P}$, together with the
auxiliary hierarchy parameter $\xi$ from \eqref{eq:xi-def}.
The dimensionless cosmic time, in particular, is defined
by $\tau\equiv t/t_\text{kick}$  with $t_\text{kick}$
from \eqref{eq:t-kick-def}. The relevant dynamic variables are
the dimensionless Hubble parameter $h(\tau)$,
the rescaled relative $q$-parameter shift $x(\tau)$,
the rescaled dimensionless energy density $r_{M1}(\tau)$
of the ultraheavy particles (called type-1),
and the rescaled dimensionless energy density $r_{M2}(\tau)$
of the massless particles (called type 2).
The type-1 particles are, for definiteness, assumed to be bosons.
In a finite temperature context without chemical potential,
the type-1 bosons have an equation-of-state parameter $w_{M1}$, which,
depends only on $M$ and $T$. The massless type-2 particles have the
equation-of-state parameter $w_{M2}=1/3$
and the effective number of degrees of
freedom $N_{\text{eff},\,2}=10^{2}$, in order to
represent the lighter particles of the Standard Model
(see Sec.~V.B of Ref.~\citen{K2010-Lambda-TeV} for further discussion).
With the temperature $T$ obtained from
the energy density $\rho_{M2}=N_{\text{eff},\,2}\,\big(\pi/30\big)\,T^4$,
it is then possible
to write $w_{M1}$ in terms of $M$ and $\rho_{M2}$ and the resulting
expression is denoted $\overline{w}_{M1}$;
see Sec.~A2 of Ref.~\citen{K2010-Lambda-TeV} for details.
Finally, we define the combination
\beq\label{eq:kappa-M1-bar-def}
\overline{\kappa}_{M1}\equiv 1- 3\,\overline{w}_{M1}\,,
\eeq
which has been found to drive the kick of the
vacuum energy density.\cite{KV2009-ew-kick}

The ordinary differential equations (ODEs) for the four dynamic variables
are\cite{K2011-Lambda-TeV-simple-model}%
\bsubeqs\label{eq:newODEs-dimensionless}
\beqa
\label{eq:newODEs-dimensionless-hdot}
\hspace*{-0mm}&&
(1-\theta_{K})\, \Big[3\dot{h} +6h^{2} -x \Big]\,
+\theta_{K}
\Big[6h\dot{h}- \dot{r}_{M1} - \dot{r}_{M2} \big]
=0,
\\[2mm]
\label{eq:newODEs-dimensionless-rM1dot}
\hspace*{-0mm}&&
\dot{r}_{M1}+(4-\overline{\kappa}_{M1})\,h\,r_{M1}
=
+g^{2}\,r_{M2}-g^{2}\,r_{M1}-\lambda^{2}\,r_{M1},
\\[2mm]
\label{eq:newODEs-dimensionless-rM2dot}
\hspace*{-0mm}&&
\dot{r}_{M2}+4\,h\,r_{M2}
=
-g^{2}\,r_{M2}+g^{2}\,r_{M1}+\lambda^{2}\,r_{M1},
\\[2mm]
\label{eq:newODEs-dimensionless-xdot}
\hspace*{-0mm}&&
(1-\theta_{K})\,
\Big[3\,h\,\dot{x}/\xi+ 3\, h^{2}\,x/\xi
-\Big(x^{2}/(2\xi) + r_{M1} + r_{M2} - 3\, h^{2}\Big)\Big]
\nonumber\\&&
+\theta_{K}\,\dot{x}=0,
\eeqa
\esubeqs
with $\theta_{K}$ given by \eqref{eq:KAnsatz-thetaK}
and $g^{2}$ by \eqref{eq:gAnsatz-all}, both in term of
the dimensionless cosmic time $\tau$
(the overdot in these ODEs denotes differentiation
with respect to $\tau$).
The source terms in Eqs.~\eqref{eq:newODEs-dimensionless-rM1dot}
and \eqref{eq:newODEs-dimensionless-rM2dot}
are somewhat different compared to those of
(3.3b) and (3.3c) in Ref.~\citen{K2011-Lambda-TeV-simple-model}.
Equations~\eqref{eq:newODEs-dimensionless-hdot}
and \eqref{eq:newODEs-dimensionless-xdot} correspond to
(3.3a) and (3.3d) of
Ref.~\citen{K2011-Lambda-TeV-simple-model} but are slightly
rewritten (actually, these rewritten ODEs were already used for the
numerics presented in Ref.~\citen{K2011-Lambda-TeV-simple-model}).
Note that \eqref{eq:newODEs-dimensionless-hdot} in the
$\theta_{K}=1$ phase (late times)
gives the derivative of the
standard Hubble equation $3h^{2}- r_{M1} - r_{M2}- r_{V}/\xi=0$ with
a term $r_{V}=(1/2)\, x^{2}=\text{const.}$,
according to \eqref{eq:newODEs-dimensionless-xdot} for $\theta_{K}=1$.

An exact solution of the ODEs \eqref{eq:newODEs-dimensionless}
for $g^{2}=0$ is given by\cite{KV2009-ew-kick}
\bsubeqs\label{eq:newODEs-exact-sol}
\beqa
\label{eq:newODEs-dimensionless-h}
h(\tau)&=&1/(2\,\tau)\,,
\\[2mm]
x(\tau)&=&0\,,
\\[2mm]
r_{M1}(\tau)&=&0\,,
\\[2mm]
r_{M2}(\tau)&=& 3\,[h(\tau)]^{2}=r_{M2}(\tau_{0})\; [a(\tau_{0})/a(\tau)]^4\,,
\eeqa
\esubeqs
where the last expression uses the scale factor $a(\tau)$,
in terms of which $h(\tau)$ is defined by $\dot{a}(\tau)/a(\tau)$.
The exact solution \eqref{eq:newODEs-exact-sol} holds
for both phases, the early one with $\theta_{K}=0$  and
the late one with $\theta_{K}=1$.
Incidentally, the dimensionless cosmic age $\tau=0.27$ corresponds
to having $T\approx M$.

The goal, now, is to study the kick of $r_V(\tau)$ by the sudden presence
of ultraheavy particles, which, in our model, results from
a nonzero coupling constant $g^{2}(\tau)$ for
$\tau \geq  \tau_{g} > \tau_\text{min}$.
For this purpose, the ODEs with $g^{2}(\tau)$ from \eqref{eq:gAnsatz-all}
are to be solved using the following boundary conditions:
\bsubeqs\label{eq:newODEs-bcs}
\beqa
\label{eq:newODEs-bc-h}
h(\tau_\text{min})&=&
1/(2\,\tau_\text{min})\,,
\\[2mm]
x(\tau_\text{min})&=&0\,,
\\[2mm]
r_{M1}(\tau_\text{min})&=&0\,,
\\[2mm]
r_{M2}(\tau_\text{min})&=& 3\,[h(\tau_\text{min})]^{2}\,,
\eeqa
\esubeqs
which matches the solution \eqref{eq:newODEs-exact-sol}
of the earlier phase $\tau < \tau_\text{min}$.

For later use, we can mention that the $\xi =\infty$
equations have been discussed in
Ref.~\citen{K2011-Lambda-TeV-simple-model} and that
an analytic result has been obtained for the
vacuum energy density without $K$--freezing effects,
\beq\label{eq:rV-xi-infty}
r_V(\tau)\,\Big|^{(\xi =\infty\,,\,\text{no\;$K$--freezing})}=
\frac{1}{8}\,
\Big[\overline{\kappa}_{M1}(\tau)\,r_{M1}(\tau)\Big]^{2}\,.
\eeq
But \eqref{eq:rV-xi-infty} still needs two $\xi =\infty$ ODEs to be solved,
in order to obtain the explicit functions $\overline{\kappa}_{M1}(\tau)$
and $r_{M1}(\tau)$.

\subsection{Dissipation model}
\label{subsec:Dissipation-model}

In this article, we primarily model the freezing of $r_V(\tau)$
by taking a particular time-dependent $K[q,t]$
and by remaining entirely within $q$--theory which
describes \emph{reversible} processes.
But, as argued in Sec.~IV of Ref.~\citen{KV2009-ew-kick}, the
freezing of $r_V(\tau)$ may very well be due to quantum-dissipative effects, 
that is, \emph{irreversible} processes.\cite{LandauLifshitz-Fluid-mechanics} 
Concretely, very low-energy gravitons (and possibly neutrinos) may be
responsible for the dissipation.\cite{ZeldovichStarobinsky1977}
Specifically, the energies of these particles must be of the order
of $\text{meV}$, which value traces back to \eqref{eq:t-kick-def}.
In any case,
a phenomenological description of this quantum dissipation has
been given in  Ref.~\citen{KV2009-ew-kick}, which we will now
briefly review.

The following model equation\cite{KV2009-ew-kick} can be used:
\begin{equation}\label{eq:rVdiss-ODE}
\dot{r}_{V}^\text{\,diss}(\tau) = - \Gamma(\tau)\,
\Big[r_{V}^\text{diss}(\tau)- r_\text{V,0}(\tau) \Big]\,,
\end{equation}
which recalls the standard description of bulk-viscosity effects in fluid
mechanics [see, in particular, Eq.~(78.1)
of Ref.~\citen{LandauLifshitz-Fluid-mechanics}].
The interpretation of \eqref{eq:rVdiss-ODE} is that
$r_\text{V,0}(\tau)$  is the ``bare'' vacuum energy density driven
by the kick from the ultraheavy particles
and that $\Gamma(\tau)\geq 0$ is the rate
at which the ``excess'' vacuum energy density is dissipated into particles.
Equation \eqref{eq:rVdiss-ODE} has an
exact solution,\cite{KV2009-ew-kick}
\beq
\label{eq:rVdiss-general-solution}
r_{V}^\text{diss}(\tau)=  \int_0^{\tau} d\tau' \;\Gamma(\tau')\, r_\text{V,0}(\tau')\;
                   \exp\left[ -\int_{\tau'}^{\tau} d\tau''\;\Gamma(\tau'')\right]\,,
\eeq
for boundary condition $\rho_{V}(0)=0$,
which holds for times $\tau$ well after the Planck era.

In the next section, we will obtain a numerical estimate
of the asymptotic value of $r_{V}^\text{diss}$
based on \eqref{eq:rVdiss-ODE}, by use of the following approximations:
\bsubeqs\label{eq:rVdiss-ODE-choices}
\beqa
\label{eq:rVdiss-ODE-choice-rV0}
r_\text{V,0}(\tau)
&=&
\frac{1}{8}\,
\Big[\overline{\kappa}_{M1}(\tau)\,r_{M1}(\tau)\Big]^{2}\,,
\\[2mm]
\label{eq:rVdiss-ODE-choice-Gamma}
\Gamma(\tau)
&=& \gamma/ \tau^{2}\,,
\eeqa
\esubeqs
where the first approximation becomes exact in the limit $\xi\to\infty$,
according to \eqref{eq:rV-xi-infty},
and the second approximation is purely illustrative.
Taking a relatively large value for $\gamma$ forces
$r_{V}^\text{diss}(\tau)$ to follow $r_\text{V,0}(\tau)$,
according to \eqref{eq:rVdiss-ODE} or \eqref{eq:rVdiss-general-solution}.
A small enough finite value of $\gamma$ allows 
$r_{V}^\text{diss}(\tau)$ to reach a nonzero asymptotic value.

Remark that \eqref{eq:rVdiss-ODE} is only part of the whole story,
as it does not specify
the detailed changes in the matter energy densities $\rho_{M,n}(\tau)$
corresponding to the change of $\rho_{V}^\text{diss}(\tau)$.
But this partial description is perfectly valid during
the kick phase,
as the amount of energy carried by the vacuum then is negligible
compared to that of the ponderable matter by a factor
of order $\xi \sim 10^{57}$,
according to \eqref{eq:xi-def} for $M \sim 10\;\text{TeV}$
(see also Fig.~1 of Ref.~\citen{ArkaniHamed-etal2000}).

\section{Numerical results}
\label{sec:Numerical-results}

\begin{figure*}[t]  
\hspace*{-0mm}  
\includegraphics[width=1\textwidth]{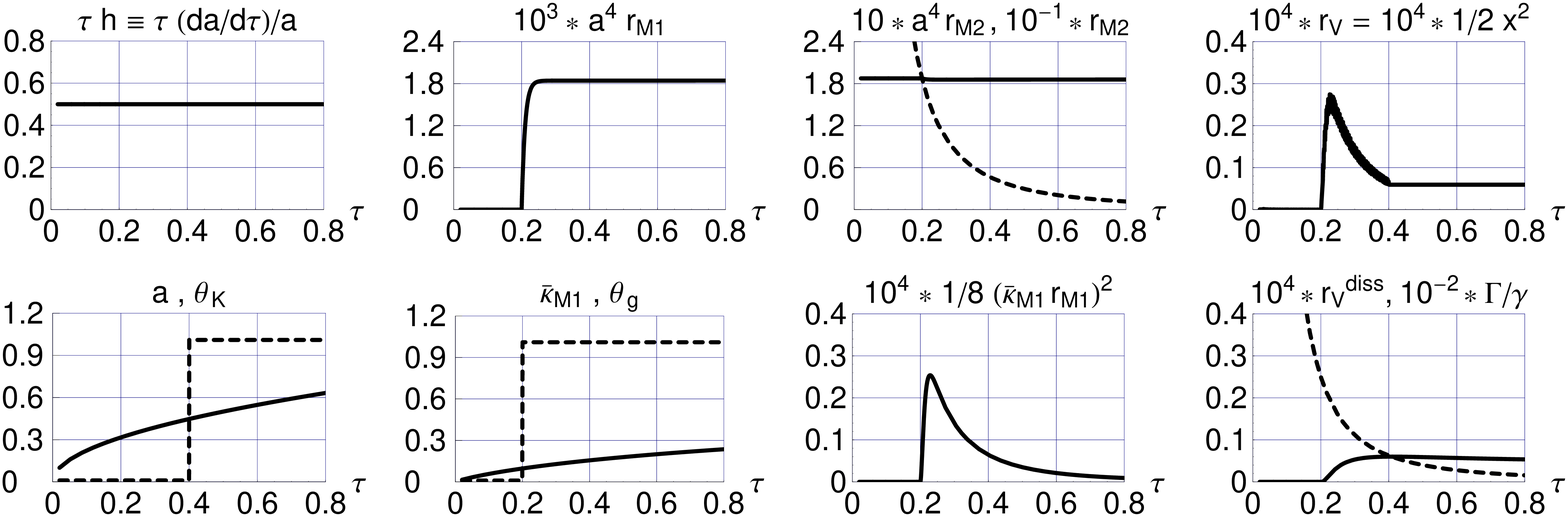}
\vspace*{-6mm}
\caption{Numerical solution of the dimensionless ODEs
\eqref{eq:newODEs-dimensionless}
with an equation-of-state function
$\overline{\kappa}_{M1}(\tau)\equiv 1- 3\,\overline{w}_{M1}(\tau)$
as defined in Sec.~A2 of Ref.~\citen{K2010-Lambda-TeV}.
The panels are organized as follows: the four basic dynamic variables
$h(\tau)$, $r_{M1}(\tau)$, $r_{M2}(\tau)$, and $x(\tau)$
are shown on the top row
and secondary or derived quantities on the bottom row.
The dashes lines in certain panels refer to the second quantity
listed in the respective panel label, for example, the dashed line in the
bottom-left panel corresponds to $\theta_{K}$.
The main result is the nonzero remnant value of the dimensionless
gravitating vacuum energy density $r_{V}\equiv x^{2}/2$
shown in the top-right panel.
The bottom-right panel shows, for comparison, the dimensionless
vacuum energy density from quantum-dissipative effects,
as modeled by \eqref{eq:rVdiss-ODE}
with approximations \eqref{eq:rVdiss-ODE-choices}.
The model parameters are
$\{\xi,\, N_{\text{eff},\,2} ,\,
\overline{g}^{2} ,\,\lambda^{2} ,\,
\tau_{g},\, \tau_{K} ,\,\gamma \}$ $=$
$\{10^7,\,10^{2} ,\,
1 ,\, 10^{2}   ,\,
0.2  ,\,  0.4 ,\,1/5 \}$
and the ultraheavy type-1 particles are assumed to be bosons
(similar results are obtained for type-1 fermions,
with somewhat lower values for $r_{V}$ and $r_{V}^\text{diss}$
by approximately $20\,\%$).
The ODEs are solved over the interval
$[\tau_\text{min},\, \tau_\text{max}]$ $=$ $[0.02,\, 0.8]$
with the following boundary conditions at $\tau=\tau_\text{min}=0.02$:
$\{x,\, h,\,       a,\,
r_{M1},\, r_{M2}\}$ $=$
$\{0,\, 25,\,     1/10,\,
0,\,     1875\}$.}
\label{fig:1}
\vspace*{0.5cm}
\includegraphics[width=1\textwidth]{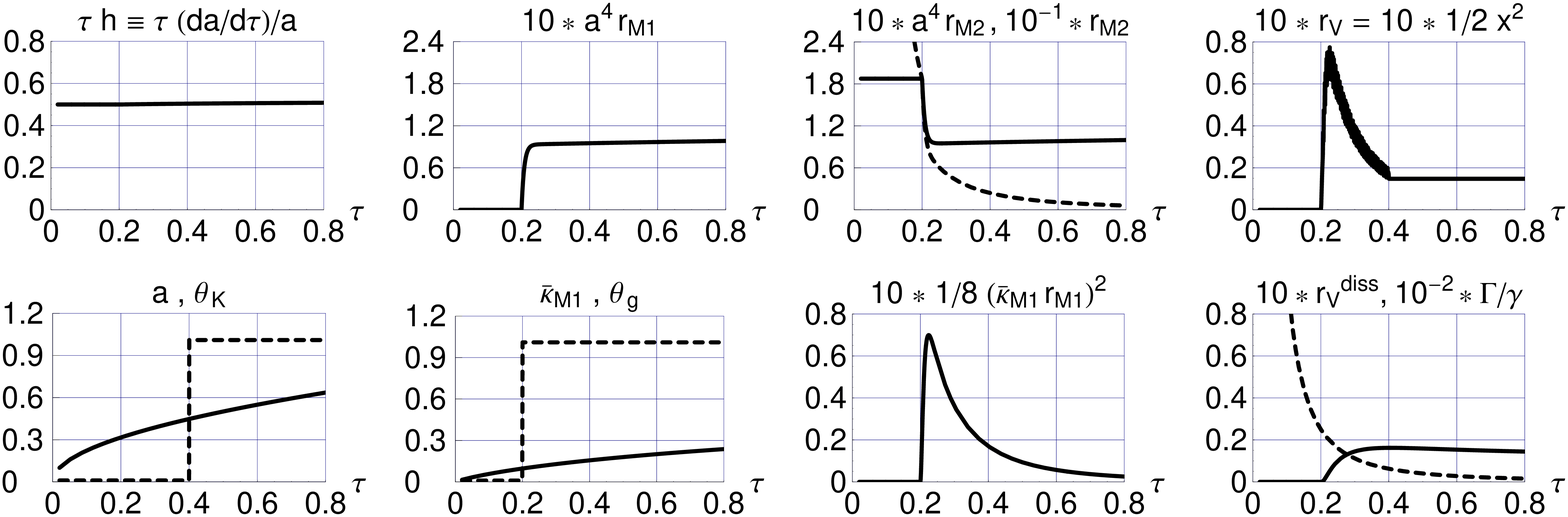}
\vspace*{-6mm}
\caption{Same as Fig.~\ref{fig:1}, but with model parameters
$\{\overline{g}^{2} ,\,\lambda^{2} \}$ $=$
$\{64   ,\, 1  \}$.
These coupling constants make for
a ratio $r_{M1}/r_{M2} \sim 1$ at $\tau=(\tau_{g}+\tau_{K})/2=0.3$,
whereas the coupling constant of Fig.~\ref{fig:1}
give a ratio of order 1/100.}
\label{fig:2}
\end{figure*}

The numerical solutions of the ODEs \eqref{eq:newODEs-dimensionless}
with boundary conditions \eqref{eq:newODEs-bcs}
are readily obtained. We present numerical solutions
in Figs.~\ref{fig:1} and \ref{fig:2} for two choices
of coupling constants $(\overline{g},\,\lambda)$,
which give a ratio $r_{M1}(0.3)/r_{M2}(0.3)$
of order $1/100$ and $1$, respectively.
In an equilibrium context, these energy-density ratios would translate
into a degrees-of-freedom ratio
$N_{\text{eff},\,1}/N_{\text{eff},\,2}=1/100$ for case 1 and
a ratio  $N_{\text{eff},\,1}/N_{\text{eff},\,2}=100/100$ for case 2.
Note that the small oscillations on $r_{V}=(1/2)\, x^{2}$
in the top-right panels of the figures disappear for
even larger values of the hierarchy parameter, $\xi \gg 10^{7}$.
At this moment, it may be useful to recall the analytic $\xi =\infty$
result \eqref{eq:rV-xi-infty}.
This analytic result suggest to consider the quantity
$1/8\,\big(\overline{\kappa}_{M1}\,r_{M1}\big)^{2}$ shown in the
third bottom-row panels of the figures, which is indeed smoother
than the numerical result $r_V(\tau)$ shown in the top-right panels.

\vspace*{0cm}

The ODEs \eqref{eq:newODEs-dimensionless} result from the $K$--freezing
model and the main result of the numerical calculation
is the asymptotic plateau of the vacuum energy density $r_V(\tau)$
as shown in the top-right panels of Figs.~\ref{fig:1} and \ref{fig:2}.
For comparison, we also show, in the
bottom-right panels, the vacuum energy density resulting
from the dissipation model \eqref{eq:rVdiss-ODE}
with approximations \eqref{eq:rVdiss-ODE-choices}
and an appropriately chosen $\gamma$ value.  
Similar results are obtained with $\Gamma(\tau)$ \textit{Ans\"{a}tze} that
drop to zero faster than $1/\tau^{2}$.
An exponential tail, for example, has been used in a previous version of the
present paper [arXiv:1503.03858v1].

For the case of Fig.~\ref{fig:1} (``$N_{\text{eff},\,1}=1\,$'')
with an $r_V$ peak of order $2 \times 10^{-5}$, the frozen
asymptotic value $r_{V\infty}$ obeys the upper bound  
\beq\label{eq:rVinfty-upperbound-case1}
r_{V\infty}\,\Big|^{\text{(case-1)}}
\leq
\max\, \big[r_{V}(\tau)\big]^{\text{(case-1)}}
\sim 10^{-5} \,,
\eeq
which translates into the following lower-bound on $M$ from
\eqref{eq:M-estimate}:
\beq\label{eq:M-lowerbound-case1}
M\,\Big|^{\text{(case-1)}} \gtrsim 10\;\text{TeV} \,.
\eeq
For the case of Fig.~\ref{fig:2} (``$N_{\text{eff},\,1}=10^{2}\,$'')
with an $r_V$  peak of order $7\times 10^{-2}$,
the asymptotic value $r_{V\infty}$ obeys the upper bound  
\beq\label{eq:rVinfty-upperbound-case2}
r_{V\infty}\,\Big|^{\text{(case-2)}}
\leq
\max\, \big[r_{V}(\tau)\big]^{\text{(case-2)}}
\sim 10^{-1} \,,
\eeq
which gives
\beq\label{eq:M-lowerbound-case2}
M\,\Big|^{\text{(case-2)}}  \gtrsim 3\;\text{TeV} \,.
\eeq

\section{Conclusion}
\label{sec:Conclusion}

The results \eqref{eq:M-lowerbound-case1} and \eqref{eq:M-lowerbound-case2}
from the $K$--freezing model of Sec.~\ref{subsec:K-freezing-model}
set the mass scale of the hypothetical new particles.
If the underlying physics is able to
relate the time scale $t_{g}$ of ultraheavy (type--1)
particle creation and the
time scale $t_{K}$ of the change in the effective gravitational
coupling, these inequalities could be replaced by rough
equalities and be all the more convincing.
Perhaps such a single physical process is similar to the
one of slow-roll particle production discussed in
the context of inflation models.\cite{KofmanLindeStarobinsky1997}
In our case, the slow-role phase must be relatively short.
More importantly, such a process must not reinstate the
cosmological constant problem (with a new scale of order $M^4$)
and must, therefore, include the evolution of $q$.
Particle creation\cite{ZeldovichStarobinsky1977} is
also crucial for quantum-dissipative effects of
the vacuum energy density, as illustrated
by the alternative model of Sec.~\ref{subsec:Dissipation-model}.

As \eqref{eq:M-lowerbound-case1} and \eqref{eq:M-lowerbound-case2}
are only lower bounds, we cannot predict that the mass scale $M$
must necessarily be in reach of the LHC
with a $13\;\text{TeV}$ center-of-mass energy
(for a possible high-luminosity upgrade, see Ref.~\citen{HiLumi-LHC}).
Perhaps a next-generation proton-proton collider with
$50-100\;\text{TeV}$ center-of-mass energy\cite{FCC,SPPC} is needed.
For the moment, we can only adopt a ``wait--and--see''
attitude.\footnote{As to the properties of the hypothetical new particles,
we remain entirely agnostic. One possibility involves ultraheavy
composite scalars from top-condensation models for the light
Higgs,\cite{Cheng-etal2013,Fukano-etal2013,VolovikZubkov2014}
but many other models exist.}
If multi--$\text{TeV}$ particles are discovered at the LHC
or a next-generation collider
(and, admittedly, this is a big `if'), an exciting prospect
may be that they provide a new window on the early universe
and fundamental physics.

\section*{Acknowledgment}

The author thanks G.E. Volovik for numerous discussions on vacuum energy
over the last years.



\end{document}